\documentclass[fleqn,10pt]{wlscirep1}
\usepackage{changes}
\usepackage{lipsum}
\usepackage{multirow}
\usepackage{hyperref}
\usepackage{fix-cm}
\usepackage{amsmath}
\usepackage{graphicx}
\usepackage{verbatim}
\usepackage{tabularx}
\usepackage{ctable}
\usepackage{array}
\usepackage{xcolor}
\usepackage{colortbl}
\usepackage{hhline}
\usepackage{xr}
\usepackage{caption}
\usepackage{graphics}
\usepackage{epsfig}

\usepackage{etoolbox}
\makeatletter
\patchcmd{\@maketitle}{ABSTRACT}{}{}{}
\makeatother

\usepackage{sidecap}

\title{Network analysis of synonymous codon usage}

\author[1,2,3]{Khalique Newaz}
\author[1]{Gabriel Wright}
\author[1,2,3]{Jacob Piland}
\author[4]{Jun Li}
\author[5]{Patricia Clark}
\author[6]{Scott Emrich}
\author[1,2,3,*]{Tijana Milenkovi\'{c}}
\affil[1]{Department of Computer Science and Engineering, University of Notre Dame, Notre Dame, IN 46556, USA}
\affil[2]{Interdisciplinary Center for Network Science and Applications, University of Notre Dame, Notre Dame, IN 46556, USA}
\affil[3]{Eck Institute for Global Health, University of Notre Dame, Notre Dame, IN 46556, USA}
\affil[4]{Department of Applied and Computational Mathematics and Statistics, University of Notre Dame, Notre Dame, IN 46556, USA}
\affil[5]{Department of Chemistry and Biochemistry, University of Notre Dame, Notre Dame, IN 46556, USA}
\affil[6]{Department of Electrical Engineering and Computer Science,  University of Tennessee, Knoxville, TN 37996, USA.}
\affil[*]{Corresponding author (email: tmilenko@nd.edu)}

\begin{abstract}
Most amino acids are encoded by multiple synonymous codons. For an amino acid, some of its synonymous codons are used much more rarely than others. Analyses of positions of such rare codons in protein sequences revealed that rare codons can impact co-translational protein folding and that positions of some rare codons are evolutionary conserved. Analyses of positions of rare codons in proteins’ 3-dimensional structures, which are richer in biochemical information than sequences alone, might further explain the role of rare codons in protein folding. We analyze a protein set recently annotated with codon usage information, considering non-redundant proteins with sufficient structural information. We model the proteins’ structures as networks and study potential differences between network positions of amino acids encoded by evolutionary conserved rare, evolutionary non-conserved rare, and commonly used codons. In 84\% of the proteins, at least one of the three codon categories occupies significantly more or less network-central positions than the other codon categories. Different protein groups showing different codon centrality trends (i.e., different types of relationships between network positions of the three codon categories) are enriched in different biological functions, implying the existence of a link between codon usage, protein folding, and protein function.
\end{abstract}
\begin{document}

\flushbottom
\maketitle

\thispagestyle{empty}

\section{Introduction}
\subsection{Motivation and related work. } The genetic code is redundant, meaning that most amino acids are encoded by more than one codon. Codons that code for the same amino acid are called synonymous codons. For an amino acid, it is usually the case that some of its synonymous codons encode it in the given genome relatively more commonly than the others \cite{sharp1986, chaney2015}. Henceforth, intuitively, when we say ``common'' codon, we mean a synonymous codon that is used frequently, and when we say ``rare'' codon, we mean a synonymous codon that is used infrequently. Precise definitions depend on which computational model is used to characterize synonymous codons as common or rare. Several such models exist \cite{liu2018}.

Rare codons are associated with lower tRNA levels, expression levels, and translational accuracy \cite{ikemura1985,sharp1987,kramer2007}. As a result, it has been hypothesized that since common codons show efficient translation, they are more likely to be under selective pressure to occupy important regions in protein structures. This has been supported by several prior efforts that have shown evolutionary conservation of optimal (i.e., common) codons in structurally important regions \cite{zhou2009, warnecke2010, pechmann2013}, while non-optimal (i.e., rare) codons tend to occur in structurally disordered regions of a protein structure \cite{zhou2015}.

Contrary to the view that rare codons are translationally inefficient and thus not optimal, several studies have argued that rare codons can actually coordinate optimal co-translational protein folding of the polypeptide chain by slowing down the translation process \cite{komar2009, kimchi2007, zhou2013, komar1999, sander2014, buhr2016,jacobson2016}. Because of this, and because folds (i.e., 3-dimensional (3D) structures) of orthologous proteins are often evolutionary conserved \cite{illergaard2009}, recent studies have hypothesized that sequence positions of important rare codons in orthologous proteins should also be evolutionary conserved \cite{chaney2017, jacobs2017}. Based on this hypothesis, Chaney et al.\cite{chaney2017} analyzed 76 fully sequenced genomes to first identify groups of orthologous proteins. Then, in the multiple sequence alignment of proteins in each orthologous group, Chaney et al.\cite{chaney2017} identified statistically significantly co-occuring (i.e., evolutionary conserved) rare codon positions. Henceforth, we denote such evolutionary consered rare codons as ``conserved rare'' codons and the rare codons that are not evolutionary conserved as ``non-conserved rare'' codons. It is difficult to show that the conservation of rare codon positions is due to positive selection because it is the usage of codons, and not the actual sequence, that is conserved across multiple sequences. However, co-occurrence of rare codon positions implies importance and hence shows that these rare codon positions are most likely involved in conserved biological functions \cite{ba2012,gonzalez2004}.

With the hypothesis that rare codons can be biochemically important,  prior efforts have also focused on finding a ``correlation'' between rare or common codons and different secondary structural elements of proteins (i.e., $\alpha$-helices or $\beta$-sheets) \cite{gupta2000, pechmann2013}. However, to the best of our knowledge, there has been no prior study that has examined structural differences between different codon categories themselves. Specifically, no previous work has asked whether conserved rare, non-conserved rare, and common codons occupy different positions in the 3D structure of a protein, which is what we ask here. 

We consider this question using the concept of networks \cite{newman2018}. That is, we model protein 3D structures as protein structure networks (PSNs). In a PSN, nodes are amino acids of the corresponding protein and there is an edge between two amino acids if they are sufficiently close in the 3D space \cite{milenkovic2009}. Network-based analysis of protein structures is promising. For example, we recently proposed two methods called GRAFENE and NETPCLASS, which used the concept of PSNs in the tasks of unsupervised and supervised \emph{global} 3D protein structural comparison, respectively \cite{faisal2017, newaz2018}. GRAFENE and NETPCLASS were shown to outperform state-of-the-art protein 3D structural [e.g., DaliLite \cite{holm2010}, TM-align \cite{zhang2005}, and GIT \cite{ harder2012}] and sequence-based [e.g., SVMfold \cite{xia2016}] methods for the same or similar purpose. In addition, several studies used PSNs to analyze the \emph{local} structure of a protein, i.e., individual nodes or groups of nodes within a PSN \cite{vendruscolo2002, amitai2004, del2006, vacic2010, brysbaert2018}. Most of such studies have focused on linking network centralities of nodes in a PSN with biochemical properties (e.g., active site involvement, ligand-binding involvement, or evolutionary conservation, but not synonymous codon usage) of the corresponding amino acids, with the goal of studying whether biochemically interesting amino acids occupy ``important'' (i.e., central) network structural positions in a PSN. 

These studies demonstrate that PSNs can accurately capture the 3D structure of a protein and indicate the promise of network-based analyses of synonymous codon usage, the task that we focus on in this paper.

\subsection{Our contributions} For the very first time, this paper hypothesizes that amino acids encoded by conserved rare, non-conserved rare, or common codons occupy different positions in the 3D structure of a protein.

To test this hypothesis, we rely on a recent study that analyzed over ${\sim}$280,000 protein sequences (technically, protein-coding nucleotide sequences) to identify co-occurring (i.e., conserved) rare codon positions within orthologous proteins \cite{chaney2017}. Given that we rely on 3D protein structural data to build PSNs, we can only focus on the subset of the ${\sim}$280,000 protein sequences that have such data in the Protein Data Bank (PDB) \cite{berman2000}. There are 16,501 such sequences. Following several additional criteria to ensure as high-quality data as possible (see Section \ref{sec:seq-pdb} for details), we filter down the 16,501 proteins down to 4,626 proteins. Of the 4,626 proteins, we can analyze only those proteins that have sufficient amount of information for each of the three codon categories in their corresponding resolved protein structure in the PDB. There are 63 such proteins. For details, see Section \ref{sec:codon-seq}.

Given the 63 proteins, we construct 63 corresponding PSNs. In order to study PSN structural positions of amino acids that are encoded by conserved rare, non-conserved rare, or common codons, we use the notion of network centrality of a node. Since different centrality measures can provide complementary information about a node's network position \cite{faisal2014}, we use six popular network centrality measures. Given a protein and a network centrality measure, we examine all possible trends between the conserved rare, non-conserved rare, or common codons. There are 24 possible trends. For example, one possible trend is that centralities of conserved rare codons are significantly greater than centralities of non-conserved rare and common codons. Another possible trend is that centralities of common codons are significantly greater than centralities of non-conserved rare and conserved rare codons. We also consider as one of the trends the scenario where the three categories of codons are indistinguishable from each other in terms of their centralities. We find that the 63 proteins show 17 different trends. Interestingly, there is no single trend that is dominant, meaning that even the most prevalent trend is shared between only 12 proteins.

We say that all proteins that show a given trend form the corresponding codon usage group. Hence, there are 17 codon usage groups.  
We examine whether proteins in a given group are more similar to each other than to proteins in the other groups. Because of our data generation process (see above), our 63 considered proteins are all sequence non-redundant to each other -- none of them are orthologous to each other as defined by Chaney et al.\cite{chaney2017}. For this reason, and because sequence information is typically well-correlated with 3D structural information, our proteins should be 3D structurally non-redundant as well, i.e., they should be structurally diverse. Hence, we do not expect proteins in a given codon usage group to be more similar to each other than to proteins in the other groups in terms of their 3D structures. Indeed, we verify this by comparing the proteins' global PSN topologies as well as their protein domain enrichments. However, because our 17 protein groups are formed solely based on codon usage trends, if the latter are related to protein folding and consequently function, we expect proteins in a given codon usage group to be more similar to each other than to proteins in the other groups in terms of their functions. Indeed, we verify this by analyzing enrichments of the codon usage groups in biological process Gene Ontology (GO) terms. We find that many of the  groups are significantly enriched (with $p$-values of $0.01$ or lower) in one or more biological process GO terms that are almost always unique to the given  group. This means  that proteins within the  groups are functionally more coherent than proteins across the groups. Thus, based on just PSN positions of the three codon categories, we are able to group structurally indistinguishable proteins into functionally distinguishable  groups.

Our results imply the existence of a link between codon usage, protein folding, and protein function.

\section{Methods}
\label{sec:data}

To examine whether sequence positions of rare codons are evolutionary conserved, Chaney et al.\cite{chaney2017} analyzed 280,436 protein sequences (techically, protein-coding nucleotide sequences) across 76 organisms. First, for each of the protein sequences, Chaney et al.\cite{chaney2017} identified clusters of rare codons using the \%MinMax algorithm \cite{clarke2008}. In more detail, an amino acid, i.e., its sequence position, is assigned a negative \%MinMax score if that amino acid and 16 of its neighbors (eight on each side) are, on average, encoded by a rare codon. Similarly, an amino acid, i.e., its sequence position, is assigned a non-negative \%MinMax score if that amino acid and its neighborhood are, on average, encoded by a more common codon. Second, Chaney et al.\cite{chaney2017} grouped the protein sequences into 18,864 orthologous groups and performed multiple sequence alignment of proteins within each orthologous group. Third, for each orthologous group, Chaney et al.\cite{chaney2017} identified regions that showed rare codon co-occurrence, i.e., regions where negative \%MinMax values were present in a statistically significantly high number of sequences in the multiple alignment corresponding to the given orthologous group. In this study, we use the same 18,864 orthologous groups and 280,436 protein sequences with the knowledge about positions of conserved rare, non-conserved rare, and common codons in multiple alignments corresponding to the orthologous groups. Specifically, we use this data as follows.

First, we map these protein sequences to their best-matching PDB sequences (Section \ref{sec:seq-pdb}). Henceforth, when we say ``protein sequence'', we mean a protein sequence that we obtain from Chaney et al.\cite{chaney2017}, and when we say ``PDB sequence'', we mean the protein sequence that we obtain from the PDB. Second, we map codon usage information from a protein sequence to the corresponding PDB sequence, in order to annotate in the latter an amino acid as either conserved rare, non-conserved rare, or common (Section \ref{sec:codon-seq}). Third, we create PSNs from 3D structures of the PDB sequences (Section \ref{sec:3d:net}). Fourth, we use the notion of network centrality  to study relationships between PSN positions of the three codon categories (i.e., conserved rare, non-conserved rare, and common codons) (Section \ref{sec:psnanalysis}). Fifth, we evaluate whether proteins showing different trends in PSN positions of the three codon categories are 3D structurally or functionally (dis)similar (Section \ref{sec:annotations}). 

\subsection{Mapping protein sequences to PDB sequences}
\label{sec:seq-pdb}
Here, we discuss key steps taken to map protein sequences to sequences in the PDB. For additional details, see Supplementary Section S1.

Since we aim to study protein 3D structures, of the 280,436 protein sequences, we wish to only keep those that can be aligned well to  sequences in the PDB, as follows.

For each of the protein sequences (recall that these are techically protein-coding nucleotide sequences), we find the corresponding best-matching PDB sequence, by aligning each protein sequence against each sequence from the PDB using BLAST \cite{altschul1990}. Specifically, as is typically done, we use BLASTX to find the best match for a query nucleotide sequence (here, one of the 280,436 sequences) in a database of amino acid sequences (here, all of the sequences in the PDB), and we use TBLASTN to find the best match for a query amino acid sequence  (here, one of the  sequences from PDB) in a database of nucleotide sequences (here, all of the 280,436 sequences). These two steps give us two sets of best-matching sequence pairs. We only keep those protein sequence-PDB sequence matches that are present in both sets, i.e., that correspond to  reciprocal best hit (RBH) pairs. We obtain 16,501 protein sequence-PDB sequence RBH pairs, where we treat the two proteins in an RBH pair as likely homologs \cite{bork1998}. 

It is possible that an RBH pair may span a larger than desirable evolutionary distance, e.g., a protein sequence from yeast might pair up with a PDB sequence from worm. To avoid this, we restrict to only those RBH pairs where in the given pair, the two proteins are from the same species. The resulting number of RBH pairs is 7,603.

One of the data post-processing filters applied by Chaney et al.\cite{chaney2017} was that for each orthologous group, any paralogs were randomly filtered out to only keep one protein sequence per species. These filtered orthologous groups were then used to identify conserved rare codon regions. Since we need to use the same conserved rare codon region data and consequently the same the same orthologous group data as generated by Chaney et al.\cite{chaney2017}, we also need to apply the corresponding data post-processing filter. The resulting number of RBH pairs is 4,682. 

Of the corresponding 4,682 PDB sequences, as is typically done, we keep only those that are at most $90\%$ sequence similar to each other, to avoid any undesirable bias that sequence redundant proteins might introduce into follow-up analyses \cite{holm1998, sikic2010}. For this, we use the cdHIT-90 dataset from the PDB, which contains 44,425 protein clusters, where in each cluster the proteins are at most $90\%$ sequence similar to each other. We select a non-redundant set of PDB sequences as follows. If any one of our PDB sequences belongs to a cdHIT-90 cluster in which none of our other PDB sequences are present, we immediately keep it. If instead it belongs to a cdHIT-90 cluster having more than one of our PDB sequences, then we keep the PDB sequence that has the highest 3D structural resolution. This results in 4,626 PDB sequences that are at most $90\%$ sequence similar to each other.

\subsection{Mapping codon usage information onto PDB structures}
\label{sec:codon-seq}
For each of the 4,626 PDB sequences, we obtain the corresponding \%MinMax values from Chaney et al.\cite{chaney2017}. Among the 4,626 PDB sequences, 440 have at least one conserved rare codon, additional 3,091 have no conserved rare codon but have at least one non-conserved rare codon, and the remaining 1,095 have no conserved or non-conserved rare codon but have only common codons. For each of the 4,626 PDB sequences, we map its \%MinMax values to the corresponding 3D structurally resolved part of the PDB sequence.

We filter the 4,626 PDB sequences such that the amount of data lost by mapping the \%MinMax values from protein sequences to their corresponding 3D structurally resolved PDB structures is as small as possible, while also retaining an as large as possible amount of PDB structural information for further analysis. Specifically, first we only keep those PDB sequences that retain at least $75\%$ of their overall \%MinMax values (i.e., common, non-conserved rare,  or conserved rare codons) in the corresponding 3D structures. Second, from this filtered set we only keep those PDB sequences that retain at least $75\%$ of negative \%MinMax values (i.e., non-conserved rare or conserved rare codons) in the corresponding 3D structures. Third, from this filtered set we only keep those PDB sequences that retain at least $75\%$ of conserved negative \%MinMax values (i.e., conserved rare codons) in the corresponding 3D structures. This results in 2,573 PDB sequences, i.e., PDB structures.

Since we want to study structural positions of amino acids encoded by conserved rare codons, non-conserved rare codons, and common codons, we can analyze only those PDB structures that have at least two amino acids in each of the three codon categories. We do this in order to ensure the absolute minimum number of amino acids in each of the three codon categories needed to meaningfully measure structural differences between them. There are 64 such PDB structures.

\subsection{Construction of PSNs from PDB structures}
\label{sec:3d:net}

For each of the 64 PDB structures, we create a protein structure network (PSN). To construct a PSN, we use the Protein Data Bank (PDB) file containing information about 3D coordinates of heavy atoms (i.e., \emph{carbon}, \emph{nitrogen}, \emph{oxygen}, and \emph{sulphur}) of amino acids of the corresponding protein. In a PSN, nodes are amino acids and there is an edge between two nodes if the corresponding amino acids are sufficiently close in the 3D space. Consequently, given a protein, its  PSN construction depends on 1) the choice of atom(s) of an amino acid to represent it as a node in the PSN and 2) a spatial distance threshold between a pair of amino acids to join them by an edge. Using four different combinations of atom choice and distance threshold choices (any heavy atom with 4 {\AA}, 5 {\AA}, or 6 {\AA}  distance thresholds, and $\alpha$-carbon with 7.5 {\AA}  distance threshold), \cite{faisal2017} showed that the choice of PSN construction strategy has a minimal effect on protein structural comparison accuracy. Hence, we consider only one of these strategies: we use any heavy atom of an amino acid as a node and join any two nodes by an edge if any heavy atoms of the corresponding two amino acids are within 4 {\AA} of each other.

Since we use the concept of network centrality to analyze the PSNs (see below), we can only consider those PSNs that are connected, i.e., that each consist of a single connected component. This is because many network centrality measures are only meaningful for a connected network. 63 of the 64 PSNs satisfy this. \textbf{The 63 PSNs are our final dataset} on which we conduct all of the following analyses. Note that the 63 PSNs, i.e., the corresponding proteins, belong to 63 orthologous groups as defined by Chaney et al.\cite{chaney2017}, i.e., each protein is in its own orthologous group.

\subsection{Protein structure network (PSN) analysis}
\label{sec:psnanalysis}

Broadly speaking, there exist two categories of network centrality measures, i.e., neighborhood-based and path-based. Intuitively, neighborhood-based centrality measures capture how large or dense the local network neighborhood of a node of interest is, while path-based measures capture how ``accessible'' the other parts of the network are from a node of interest. Since the two categories of centrality measures (and even different measures within the same category) capture at least somewhat complementary network structural information, we use the three popular neighborhood-based measures and three popular path-based measures to capture positions of nodes in PSNs \cite{faisal2014}.

The neighborhood-based measures are \emph{degree centrality (DEGC)},  \emph{graphlet degree centrality (GDC)}, and \emph{k-coreness centrality (KC)}. DEGC measures the number of direct network partners of a node. GDC measures how large and dense the \emph{extended}  neighborhood of a node is: it uses the concept of graphlets (small subgraphs) and computes a weighted sum of all 2-5-node graphlets in which a given node participates \cite{milenkovic2011}. KC measures the maximum $k$-core that a node participates in, where a $k$-core of a network is the maximal subset of nodes such that each node in the subset is connected to at least $k$ other nodes in the subset. 

The path-based measures are \emph{closeness centrality (CLOSEC)}, \emph{eccentricity centrality (ECC)}, and \emph{betweenness centrality (BC)}. CLOSEC measures how close a node is on average to all other nodes in a network, by computing the inverse of the sum of shortest paths from a given node to all of the other nodes. ECC is similar to CLOSEC, as it measures how close a node is to the furthest node; the difference between ECC and CLOSEC is that ECC measures the inverse of the maximum (instead of the sum in case of the CLOSEC) of the shortest paths from a given node to all of the other nodes in a network. BC measures the percentage of all shortest paths in a network that pass through the node of interest.

Independent of whether one is comparing centrality values of different nodes in a single network or centrality values across different networks, it is typically the ranking of the nodes in terms of their centrality values that is taken into account, and not the nodes' raw centrality values. Since we compare centrality values of conserved rare, non-conserved rare and common codons in each individual PSN, and since we also compare codon centrality trends (i.e., types of relationships between network positions of the three codon categories) across the different PSNs, we also account for the ranking of nodes in a PSN, i.e., the nodes' normalized rather than raw centrality values. Specifically, given a PSN and a network centrality measure: 1) We compute raw centrality values of all nodes in the PSN. 2) We rank the nodes from the most central to the least central in increments of 0.1\%. That is, we take the maximum and minimum centrality values and define 1,000 equal-size intervals (i.e., bins) between them. 3) We assign to each node as its rank, i.e., normalized centrality value, the value of the bin to which the given node's raw centrality belongs. For example, say that for a PSN and DEGC, the maximum and minimum centrality values are 51 and 1, respectively. Then, we would define 1,000 bins between the centrality values of 51 and 1, where each bin is of size $(51-1)/1000=0.05$. Next, we would assign rank of 1 to all nodes that have DEGC values in the first bin (between 51 and 50.95), a rank of 2 to all nodes that have DEGC values in the second bin (between 50.95 and 50.90), and so on.
Importantly, note that we have checked the effect of using our normalized centrality values instead of using raw centrality values, by redoing our entire analysis with the latter. We have found only a negligible difference between the two sets of results (Supplementary Section S2).

\subsection{Structural or functional (dis)similarities between proteins in different codon usage groups}
\label{sec:annotations}

To study whether proteins in a given codon usage group are more similar to each other than they are to proteins in the other groups, we: 1) compare 3D structures of all proteins via GRAFENE \cite{faisal2017}, 2) compute enrichments of all codon usage groups in  structural domain categories from CATH and SCOP databases \cite{greene2006cath, murzin1995scop}, and 3) compute enrichments of  all groups in biological process Gene Ontology (GO) terms, relying on the GO annotation data from the PDB \cite{berman2000}.

GRAFENE works as follows. Given a PSN (i.e., protein), GRAFENE extracts a network-based feature of the protein, which counts how many graphlets (subgraphs) of each type, such as triangles or squares, appear in the PSN \cite{prvzulj2004}. Then, to compute 3D structural similarity between two PSNs, GRAFENE compares their graphlet features. We use GRAFENE to compute  similarities between all of our considered proteins. Then, we evaluate whether similarities within a given codon usage group are significantly different ($p$-value $\leq$ 0.01) than similarities between that group and all other groups, with respect to the Wilcoxon signed-rank test.

We compute enrichment of a codon usage group in a 3D structural domain category or biological process GO term as follows. Since the notion of enrichment of a group with only one to two proteins is not well defined, we  consider only groups that have more than two proteins. 13 of our 17 codon usage groups (Section \ref{sec:num-of-proteins}) satisfy  this condition, and the 13 groups cover 58 out of all 63 proteins. Additionally, we consider only structural domain categories and GO terms that annotate at least two proteins in at least one of the 13 groups; otherwise,  by definition, no enrichment can be observed in any group. Note that barely any structural domain categories from the second, third, or fourth level of CATH and SCOP hierarchies satisfy this, which is why we only consider first-level categories.  This results in seven structural domain categories (over both CATH and SCOP) and 21  GO terms. As is typically done, we use the hypergeometric test \cite{falcon2008} to compute the likelihood of obtaining the given enrichment by chance. Intuitively, given the observed number of occurrences of a structural domain category or GO term in a group of the given size, the hypergeometric test measures the probability of getting the same or higher number of occurrences of the same structural domain category or  GO term in a randomly chosen protein group of the same size. The latter is selected by chance from a set of background proteins; as this set, we use all proteins in which each protein has at least one conserved rare codon (Section \ref{sec:codon-seq}). We say that a structural domain category or GO term is statistically significantly enriched in a codon usage group if its enrichment $p$-value is $\leq 0.01$. 

Note that our data is small in terms of the number of proteins (63) and the sizes of the codon usage groups in which we measure enrichment (the largest of which has 12 proteins; Section \ref{sec:num-of-proteins}). Because of this, we are unable to  perform a multiple hypothesis correction, because by default such a correction would make any $p$-value non-significant simply because of the small data size \cite{feise2002}.

Also, note that a protein can have more than one structural domain of a given type (e.g., it can have $k$ $\alpha$-helices). We can count such multiple occurrences of a domain in a protein in two ways: we can either treat the multiple occurrences as a single occurrence (the count of $\alpha$-helices in the given codon usage group is increased by $1$), or we can treat them as multiple occurrences (e.g., the count of $\alpha$-helices in the group is increased by $k$). We have verified that this choice has no effect on the domain enrichment results. Hence, for simplicity, we  focus only on the former way.

\section{Results and discussion}
In each of the 63 PSNs, we study differences between structural positions (i.e., network centrality values) of conserved rare, non-conserved rare, and common codons, i.e., we aim to identify different trends between network positions of the three codon categories (Section \ref{sec:num-of-proteins}). As a control, we argue that it should not be possible to observe any structural differences between the three codon categories when using randomized codon usage data  (Section \ref{sec:rand-proteins}) or solely because of protein (i.e., PSN) size (Section \ref{sec:size-proteins}). As a way to validate groups of proteins corresponding to the different codon usage trends (observed from the actual data), we examine whether proteins in the given group are more 3D structurally or functionally similar to each other than to proteins in the other groups (Sections \ref{sec:structure} and \ref{sec:function}). We do not expect the former, since our proteins are sequence and thus likely 3D structurally non-redundant. We hope for the latter, because this would imply that our  structurally diverse proteins can be grouped into functionally coherent groups based solely on  codon usage information.

\subsection{Different protein groups show different trends between network structural positions of conserved rare, non-conserved rare, and common codons}
\label{sec:num-of-proteins}

Given a PSN, we use six network centrality measures to evaluate whether the three categories of codons occupy different PSN positions. Namely, given a centrality measure and a PSN, we compute centralities of all nodes in the PSN (Section \ref{sec:data}). Then, we test whether centralities of all nodes from one of the three codon categories are significantly larger than centralities of all nodes from each of the other two codon categories (Supplementary Fig. S1); we test the significance using the Wilcoxon signed-rank test.

\begin{figure}[!h]
	\centering
	\includegraphics[scale=0.29]{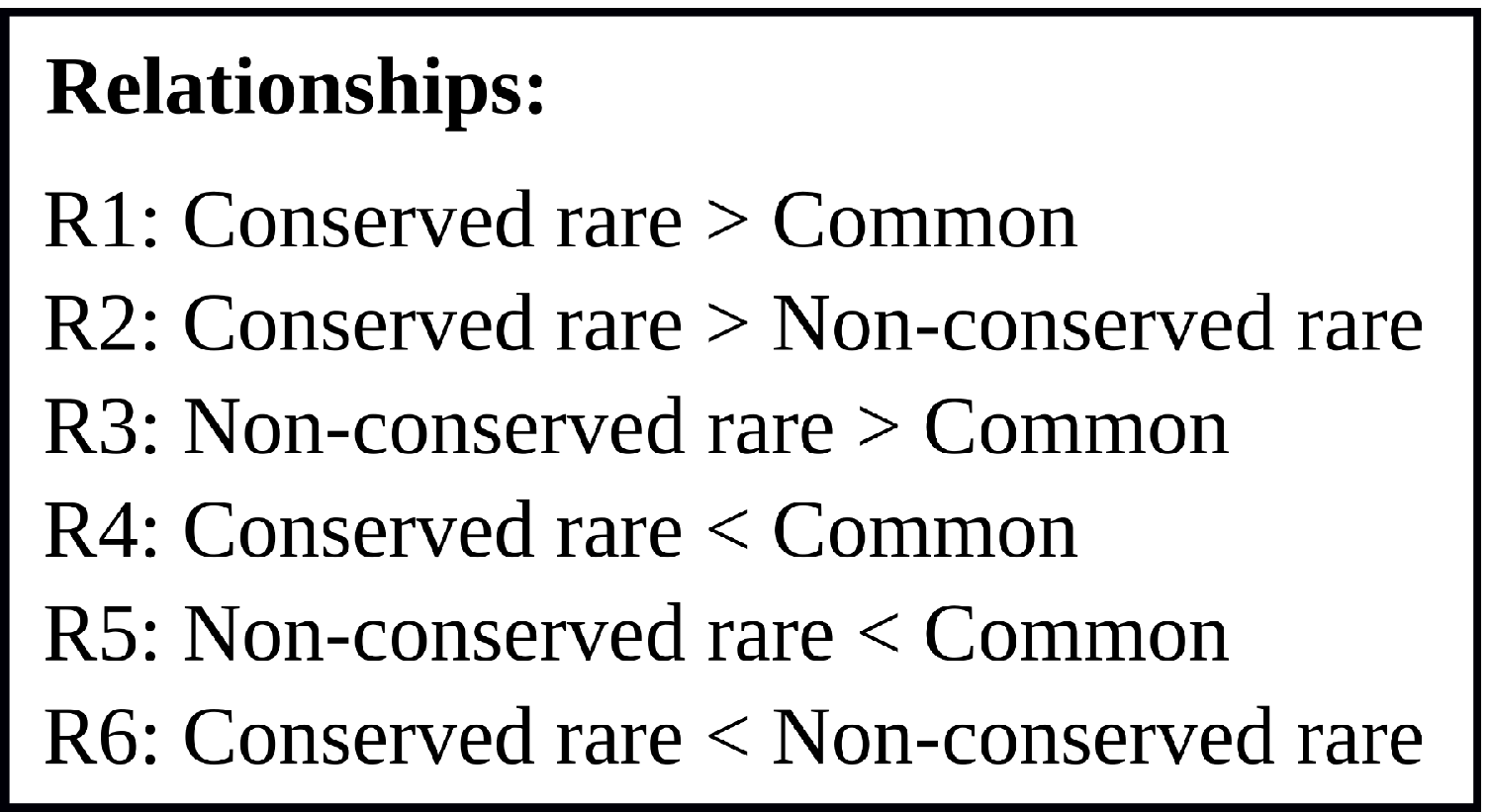}
    \caption{The six possible relationships between amino acids in a protein (i.e., nodes in a PSN) encoded by conserved rare, non-conserved rare, and common codons.}
    \label{fig:fig1}
\end{figure}

Specifically, given a network centrality measure and a PSN, we check whether centralities of conserved rare codons are significantly larger than centralities of each of non-conserved rare codons and common codons. Consequently, for  conserved rare codons, we perform two pairwise comparisons: one against non-conserved rare codons, and the other one against common codons. In the similar manner, we perform two pairwise comparisons for non-conserved rare codons: one against conserved rare codons, and the other one against common codons. Finally, we perform two pairwise comparisons for common codons as well: one against conserved rare codons, and the other one against non-conserved rare codons. Hence, we perform a total of six pairwise comparisons. We refer to the six types of comparisons as six corresponding relationships, as defined in Fig. \ref{fig:fig1}. For example, the comparison checking whether conserved rare codons have significantly larger centralities than common codons corresponds to relationship $R1$ in Fig. \ref{fig:fig1}.

We perform the above six comparisons (i.e., test the six relationships) for each of the 63 proteins (i.e., PSNs), using each of the six network centrality measures. Hence, we perform $6\times63\times6=$2,268 comparisons, i.e., Wilcoxon signed-rank tests, and  obtain 2,268  corresponding $p$-values. We apply False Discovery Rate (FDR) correction \cite{benjamini1995} to the 2,268 $p$-values and obtain the corresponding adjusted $p$-values. We identify a comparison test as positive (i.e., the corresponding relationship as significant) if its adjusted $p$-value is less than or equal to $0.05$.

We say that \emph{a protein exhibits a relationship} if the relationship is significant in the protein with respect to \emph{at least one centrality measure}. Different centrality measures can be complementary, meaning that a protein may exhibit a relationship with respect to some centrality measure(s) but not with respect to other(s), or that a protein may exhibit different relationships with respect to different centrality measures. We believe that as long as at least one centrality measure detects a relationship as significant in the given protein, the protein should be considered as exhibiting that relationship.

We find that 53 out of all 63 proteins exhibit at least one of the six relationships. In other words, 10 out of the 63 proteins exhibit none of the six relationships for any of the centrality measures. That is, PSN positions of conserved rare, non-conserved rare, and common codons are indistinguishable from each other.

\begin{figure}[!h]
	\centering
	\includegraphics[scale=0.3]{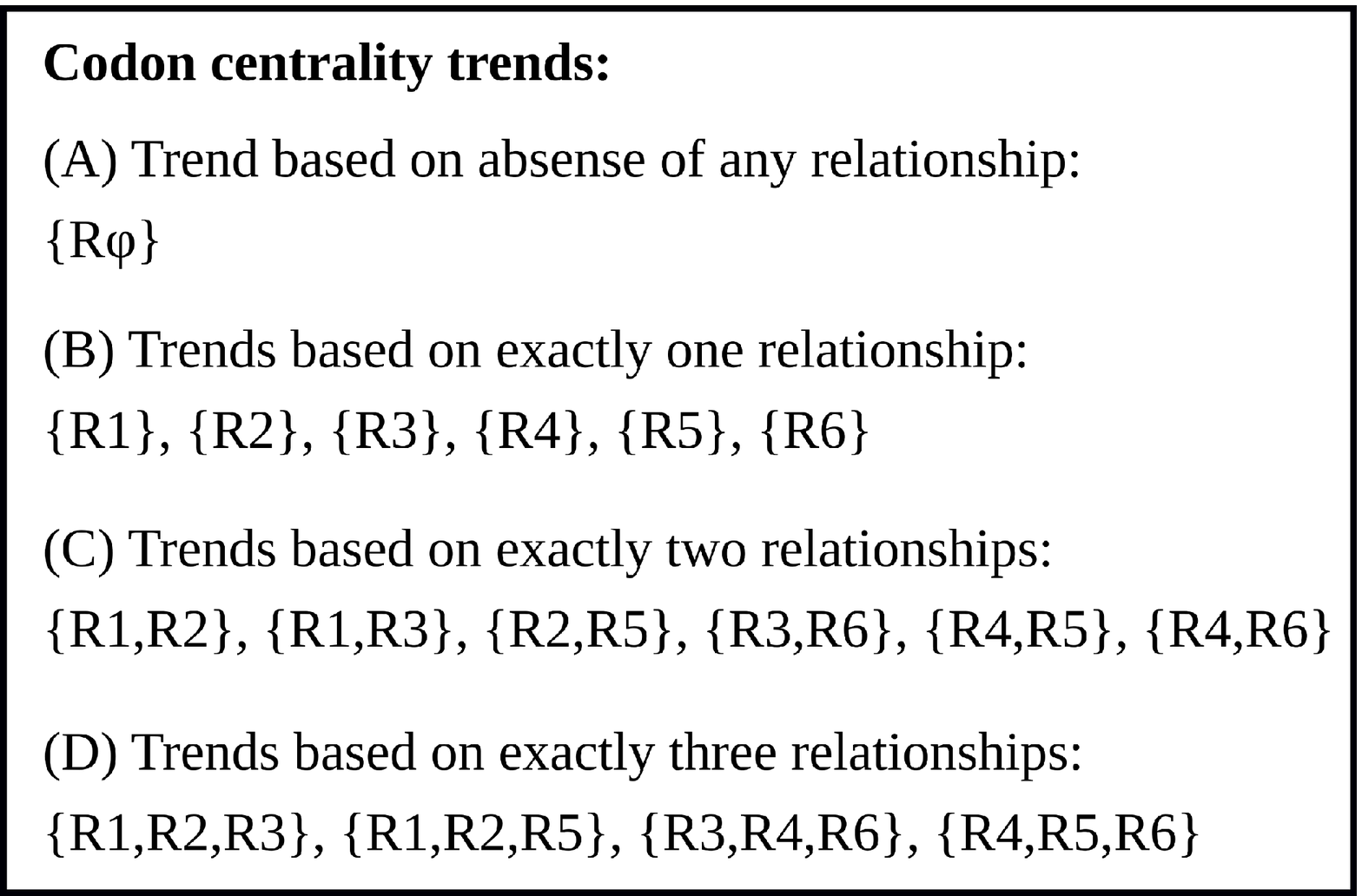}
    \caption{The 17 different codon centrality trends (i.e., different combinations of relationships between PSN positions of the three codon categories) present in our data.}
    \label{fig:fig2}
\end{figure}

Of the other 53 proteins, a protein can exhibit exactly one, exactly two, or exactly three of the six relationships from Fig. \ref{fig:fig1}. We refer to such combinations of relationships as \emph{codon centrality trends} (Fig. \ref{fig:fig2}). There are six possible trends based on exactly one relationship, all of which exist in our data.  Additionally, there are 12 possible trends based on exactly two relationships, six of which exist in our data. For example, a PSN can exhibit relationship $R1$ from Fig. \ref{fig:fig1}, i.e., conserved rare codons being more network-central than common codons, and the same PSN can also exhibit relationship $R2$ from Fig. \ref{fig:fig1}, i.e., conserved rare codons being more network-central than non-conserved rare codons. In this case, the PSN in question would exhibit trend $\{R1,R2\}$ from Fig. \ref{fig:fig2}. Finally, there are six possible trends based on exactly three relationships, four of which exist in our data. For example, a PSN can exhibit each of $R1$, $R2$, and $R3$ relationships from Fig. \ref{fig:fig1} and would thus exhibit trend $\{R1,R2,R3\}$ from Fig. \ref{fig:fig2}.

In addition, we treat the 10 PSNs that exhibit none of the six relationships (see above) as an additional ``no codon usage'' trend, referred to as $\{R\varphi\}$ in Fig.  \ref{fig:fig2}. 
Thus, in total, we observe in our data $6+6+4+1=17$ different trends. We say that all proteins that exhibit a given trend form the corresponding codon usage group. So, we have 17 codon usage groups. Henceforth, we use the terms ``trend'' and ``group'' interchangeably.

The sizes of the different groups are shown in Fig. \ref{fig:fig4} (and for informational purpose, their per-species memberships are shown in Supplementary Table S1). Interestingly, none of the 17 codon usage trends is dominant, meaning that even the largest of the groups (i.e., $\{R5\}$ and $\{R1,R2\}$) each contain ``only'' 12  out of all 63 proteins. In other words, with the exception of several very small groups, a majority of the groups are of relatively similar size and are thus well representative of the data. Hence, all of them could be biochemically interesting. We analyze the different groups from several aspects in the following sections.

\begin{figure}[!h]
	\centering
	\includegraphics[scale=0.7]{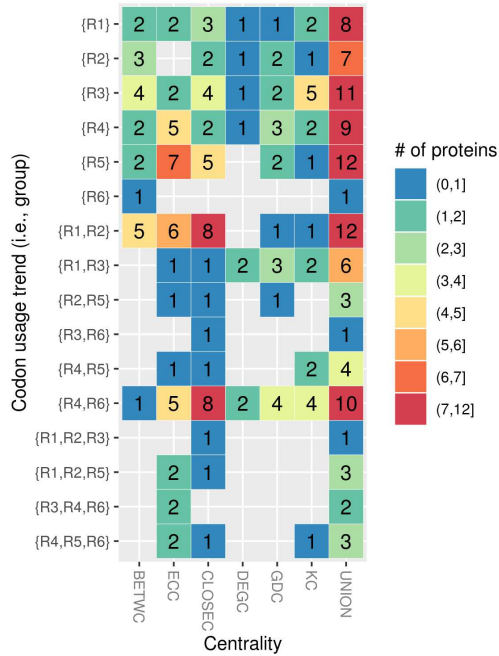}
    \caption{Numbers of proteins having the different codon usage trends. The 16 trends (i.e., codon usage groups) that exhibit at least one relationship with respect to at least one centrality measure are shown. The 17$^{th}$ ``no codon usage'' group with 10 proteins is left out, since no relationship is exhibited with respect to any centrality measure. The figure can be interpreted as follows. As an illustration, there are five proteins that have trend $\{R1, R2\}$ with respect to BETWC. Because a protein can show a trend with respect to multiple centralities, the same protein can be present in the multiple columns of the  row corresponding to that trend. So, the  ``UNION'' column shows the total number of \emph{unique} proteins in a given row, denoting the number of proteins showing the corresponding trend with respect to \emph{at least one of the six centrality measures}. For example, there are 12 unique proteins that exhibit trend $\{R1,R2\}$ with respect to at least one centrality measure (of which CLOSEC covers eight on its own). Also, because a protein can show one trend with respect to one centrality measure and another trend with respect to another centrality measure, the same protein can be present in multiple corresponding rows of the UNION column (Supplementary Fig. S2). 34 of the 63 proteins show exactly one trend over all centrality measures and are thus present in a single row, while the remaining 19, 9, and 1 proteins show exactly two, three, and four trends, respectively. Another observation is that with the exception of trends/groups of size 1, no single centrality measure captures all of the proteins in a given group, which confirms that the different centrality measures are at least somewhat complementary. }
    \label{fig:fig4}
\end{figure}

\subsection{No meaningful codon usage trends can be observed from randomized codon usage data}
\label{sec:rand-proteins}

If there is some biochemical signal behind our identified codon usage groups, we expect that if we randomize the codon usage data (i.e., randomly reshuffle labels corresponding to conserved rare, non-conserved rare, and common codons in each PSN) and then repeat the exact same analysis as above on the randomized data, (almost) none of the 63 proteins will show any significant difference between centralities of conserved rare, non-conserved rare, and common codons, i.e., (almost) all of them will belong to the $\{R{\varphi}\}$ codon usage group.  Indeed, when we randomize the codon usage data 1,000 times to account for the non-deterministic nature of the randomization procedure, we find that $62.96$ proteins belong to this group on average over the 1,000 runs, with the corresponding standard deviation of $0.2$ (Supplementary Fig. S3).

\subsection{Different codon usage groups do not show size differences}
\label{sec:size-proteins}
If there is some biochemical signal behind our identified codon usage groups, the  groups should not be biased by PSN (i.e., protein) sizes. That is, no group should have proteins that are significantly smaller or larger than proteins in the other groups. Specifically, a significant difference would imply that proteins in the different codon usage groups have different sizes and that the different trends between non-conserved rare, conserved rare, and common codons are likely ``correlated'' with protein sizes. However, a non-significant difference would imply that the different trends between non-conserved rare, conserved rare, and common codons cannot be dependent on protein sizes. To evaluate this, we compare protein sizes in each of the codon usage groups to protein sizes in all other groups using the Wilcoxon signed-rank test. By ``size'' of a protein, i.e., its PSN, we mean either the number of nodes in the PSN (i.e., the length of the protein chain) or network density of (i.e., the percentage of all possible edges that exist in) the PSN; we test both.  We find no significant differences, as all adjusted $p$-values range from $0.44$ to $0.91$ (Supplementary Fig. S4).


\subsection{All proteins are structurally diverse and thus different codon usage groups do not show structural differences}
\label{sec:structure}
By the design of our study, all of the 63 considered proteins are sequence non-redundant (i.e., at most 90\% sequence similar) to each other. Additionally, the 63 proteins belong to 63 orthologous groups as identified by Chaney et al.\cite{chaney2017} (Supplementary Fig. S5). For these reasons, and because sequence information is typically well-correlated with 3D structural information, the 63 proteins should be 3D structurally non-redundant as well. To confirm this, we evaluate protein structural (dis)similarities within and across the codon usage groups in two  ways.

First, we evaluate whether structures of PSNs (i.e., proteins) within each codon usage group are more similar than structures of PSNs across the groups by comparing network patterns of the PSNs using GRAFENE (Section \ref{sec:annotations}). Indeed, we find none of the evaluation tests to be statistically significant   (Supplementary Fig. S6). That is, proteins within a group are just as dissimilar as proteins across the groups in terms of network topology of the corresponding PSNs. 

Second, we evaluate whether different codon usage groups are enriched in different structural domain categories from CATH and SCOP (Section \ref{sec:annotations}). Indeed, we find that none of the groups show significant enrichment in any of the structural domain categories (Supplementary Fig. S7).

In summary, both analyses confirm that the 63 proteins are structurally  diverse, which is why the different codon usage group are structurally indistinguishable from each other. It is especially because of this why it would be encouraging if the different codon usage groups are enriched in different biological functions. If so, this would mean that the structurally indistinguishable proteins can be grouped into functionally coherent and distinguishable groups  based solely on codon usage information. So, next we examine potential functional enrichment of the groups.



\begin{figure*}[!b]
	\centering
	\includegraphics[scale=0.48]{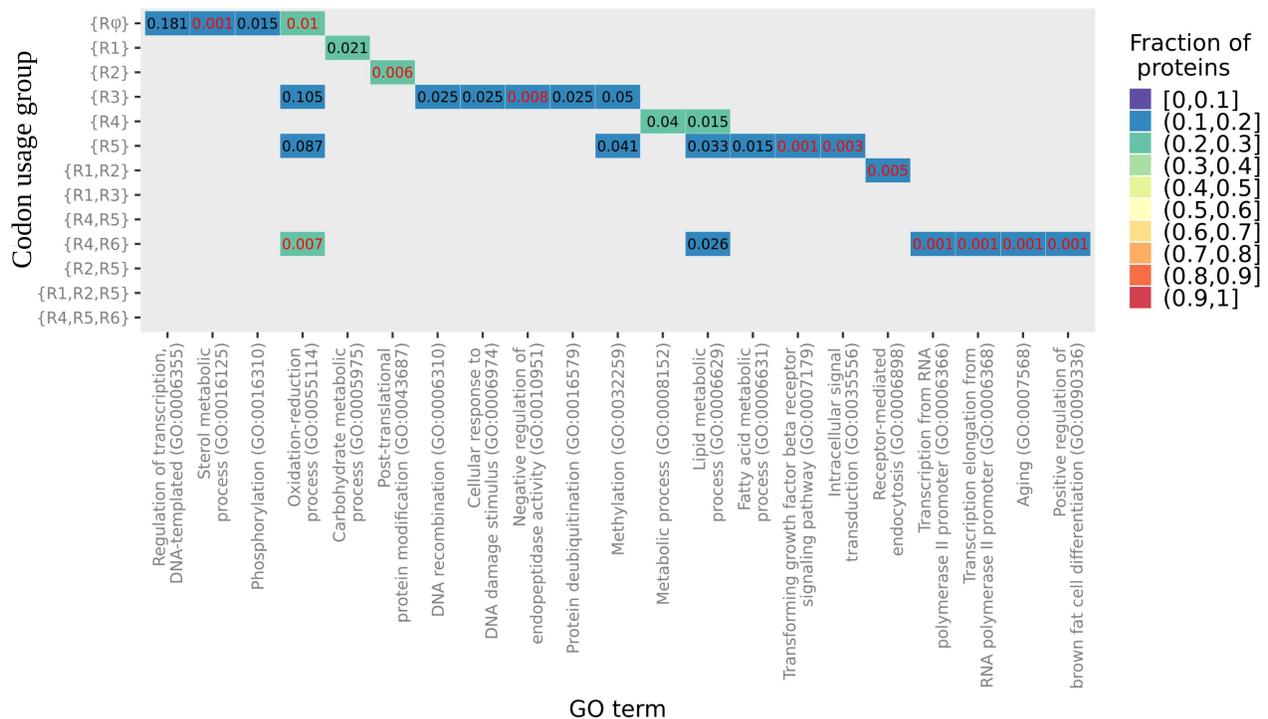}
    \caption{Functional enrichment of codon usage groups in terms of biological process GO terms. We consider those 13 out of all 17 groups that have more than two proteins, and we consider only those biological process GO terms that annotate at least two proteins in at least one of the 13 groups (Section \ref{sec:num-of-proteins}). In the figure, a colored matrix cell indicates that the given GO term annotates at least two proteins in the given  group, the color of the cell indicates the fraction of proteins in the group that are annotated by the  GO term, and the number inside the cell represents the $p$-value of the enrichment of the  GO term in the  group. The significant ($\leq 0.01$) $p$-values are colored in red. }
    \label{fig:fig3}
\end{figure*}

\subsection{Different codon usage groups \emph{do} show functional differences}
\label{sec:function}
Since our protein groups are formed solely based on codon usage trends, if the latter are related to protein folding and consequently function, proteins in a given codon usage group should be more similar to each other than to proteins in the other codon usage groups in terms of their functions. To evaluate whether this holds, we measure enrichment of each codon usage group in each GO term (Section \ref{sec:annotations}). Indeed, we find that many groups are significantly enriched in at least one GO term (Fig. \ref{fig:fig3}). Importantly, when a GO term is significantly enriched in some group, the same GO term is almost never enriched in any other group. Specifically, of all 11 GO terms that are significantly enriched in at least one codon usage group, 10 are enriched in exactly one group, and only one is enriched in more than one group. That is, proteins in the different groups are involved in different biological functions (Fig. \ref{fig:fig3}).

\section{Conclusion}
We use the concept of networks to compare 3D structural positions of amino acids encoded by conserved rare, non-conserved rare, and common codons, i.e., to analyze codon usage trends in our considered 63 proteins. We find that the 63 proteins show 17 different codon usage trends, i.e., form 17 corresponding codon usage groups. We show that even though the 63 proteins are structurally diverse and thus the codon usage groups are structurally indistinguishable from each other, the 17 protein groups are functionally distinguishable, i.e., they are enriched in different biological functions. Because of this, and because the protein groups are formed solely based on the codon usage information, we conclude that there is likely a functional ``bias'' in how the conserved rare, non-conserved rare, and common codons are positioned in the 3-dimensional structure of a protein. 

\section*{Funding}
This work is funded by a grant from the National Institutes of Health (R01 GM120733).

\bibliography{document}

\clearpage



\end{document}